\documentclass[12pt]{article}
\usepackage{graphicx}
\usepackage{float}
\usepackage{amsfonts}

\font\openface=msbm10 at10pt
\def\Integers      {{\hbox{\openface Z}}}

\def\Reals         {{\hbox{\openface R}}}
\def\Complexes     {{\hbox{\openface C}}}

\def\eight{ 8 }  

\def\tr{\mathop {{\rm \, Tr}} \nolimits}	 

\def\trace{\tr}

\def\span{\mathop {\rm span }\nolimits}

\def\implies{\Rightarrow}

\def\sqr#1#2{\vcenter{
  \hrule height.#2pt 
  \hbox{\vrule width.#2pt height#1pt 
        \kern#1pt 
        \vrule width.#2pt}
  \hrule height.#2pt}}

\def\lto{\mathop
        {\hbox{${\lower3.8pt\hbox{$<$}}\atop{\raise0.2pt\hbox{$\sim$}}$}}}
\def\gto{\mathop
        {\hbox{${\lower3.8pt\hbox{$>$}}\atop{\raise0.2pt\hbox{$\sim$}}$}}}
\def\fourth{{1 \over 4}}

\def\part{\subseteq}		

\def\ptl{\partial}

\def\braces#1{ \{ #1 \} }

\def\to{\mathop\rightarrow}	

\def\ideq{\equiv}		

\def\interior #1 {  \buildrel\circ\over  #1}     

\def\basisvector#1#2#3{
 \lower6pt\hbox{
  ${\buildrel{\displaystyle #1}\over{\scriptscriptstyle(#2)}}$}^#3}

\def\eps{\varepsilon}

\def\tilde{\widetilde}		
\def\hat{\widehat}		

\def\linebreak{\hfil\break}

\def\1{{\bf 1}}
\def\Omtilde{{\tilde\Omega}}
\def\partialderiv#1#2#3{{\left({\partial #1 \over \partial #2}\right)_#3}}

\def\L{{\cal L}}

\def\E{{\mathfrak E}}
\def\M{{\mathfrak M}}
\def\J{{\mathfrak J}}
\def\Q{{\mathfrak Q}}

\def\qplus{q_{+}}
\def\qminus{q_{-}}
\def\betahat{{\hat\beta}}
\def\Omegahat{{\hat\Omega}}

\def\Z{{\Integers}}
\def\th{\theta}
\def\diag{{\rm diag}}

\def\text#1{{\rm #1}}

\def\ave#1{\langle #1 \rangle}

\def\journaldata#1#2#3#4{\textit{#1} \textbf{#2,} #3 (#4)}

\def\span{\mathop {\rm span }\nolimits}
\def\eprint#1{{\tt #1}}

\def\REMARK{\noindent {\csmc Remark \ }}
\def\REMARK{\medskip\noindent {\bf Remark \ }}

\begin{document}
\bigskip 
\begin{titlepage}   
\bigskip \begin{flushright}   

hep-th/0703030\\ 

\end{flushright}   

\vspace{1cm}   
   
\begin{center}   
{\Large \bf 
 {The disjointed thermodynamics of rotating black holes with a NUT twist}}\\ 
\end{center} 
\vspace{2cm} 
\begin{center} 
 A.M. Ghezelbash,{%
\footnote{%
EMail: amasoud@avatar.uwaterloo.ca}}  
R.B. Mann,\footnote{ 
EMail: mann@avatar.uwaterloo.ca}\\ 
Department of Physics and Astronomy, University of Waterloo, \\ 
Waterloo, Ontario N2L 3G1, Canada\\ 
\vspace{10pt} 
Rafael D. Sorkin\footnote{ 
EMail: rsorkin@perimeterinstitute.ca}\\ 
Perimeter Institute for Theoretical Physics\\ 
31 Caroline Street North \\ 
Waterloo, ON  N2L 2Y5, Canada\\ 
and\\
Department of Physics, Syracuse University\\
Syracuse, NY 13244-1130, U.S.A.\\
\vspace{1cm} 

\end{center} 
 
\begin{abstract} 
\end{abstract} 
We study the solutions of the 
Euclidean-signature Einstein equations
(gravitational instantons) 
whose line-element takes the Kerr-bolt form, 
characterized by three real parameters:
a size, a NUT charge, and a spin rate.
The exclusion of singularities
eliminates most combinations of these parameters,
leaving only separated solution-manifolds 
between which continuous transitions are impossible
(The angular velocity divided by the temperature is forced to be a
rational multiple of $2\pi$).
This  ``quantization''  
prevents 
the free variations 
presupposed by an equation like
$T dS=dM + \Omega dJ$,
%
%
and thereby renders the first law
true, 
false,
meaningless, 
or
tautological,
depending on how one approaches it.
\end{titlepage}\onecolumn

\begin{center}
\bigskip
\end{center}

\bigskip 

\section{Introduction}

Black hole thermodynamics has been a subject of intense study for more than
30 years. It has provided us with crucial insights into quantum gravity,
and has raised some basic questions concerning quantum theory. 
Most of this research has concentrated on
spacetimes whose topology at infinity is a product of some spatial
manifold $\Sigma $\ with time, which in the (Euclidean-signature)
thermodynamic path integral becomes 
$\Sigma \times S^{1}$.

In recent years spacetimes with NUT charge and their applications have been
studied extensively \cite{DM}--\cite{Chamblin}. 
Such spacetimes differ from ordinary asymptotically flat
spacetimes because, near infinity, their topology is that of a
circle fibred non-trivially over the spatial manifold. 
In the 4-dimensional case  
the topology at infinity has most often been
the Hopf-fibration of the 3-sphere
or one of its quotients (the spatial topology 
being maintained as that of a 2-sphere).

In spite of these topological complications, 
the path-integral
formulation of gravitational thermodynamics has been carried through in
some cases, although, as we will see in detail below, 
the calculations are subject to ambiguities.
Without fully resolving these ambiguities, one has computed conserved
quantities and gravitational entropy.  
Unlike when the topology is that of a product, the entropy of NUT-charged
spacetimes has not turned out to be proportional to the area of the
event horizon.
%
%
NUT-charged spacetimes have also provided interesting tests of the
AdS-CFT correspondence and have also furnished counterexamples to
conjectured maximal masses and maximal entropies of
asymptotically de Sitter spacetimes \cite{CGM1, CGM2}~.
In addition, these metrics have played an important role in the
Kaluza-Klein setting, where their nontrivial bundle structure
corresponds to nonzero magnetic charge \cite{KKmonopole}.

Comparatively little is known about the properties of NUT-charged
spacetimes with nonzero rotation.  These ``Kerr-bolt'' spacetimes are exact
solutions to the Einstein equations, and are known both with and
without a cosmological constant.  Although a computation of their
conserved mass and angular momentum 
was carried out several years ago
\cite{robb}
(with their
 entropy being inferred from the relation (\ref{GDrel}) whose
 derivation we recapitulate below),
there has been no investigation of the gravitational
thermodynamics of this class of spacetimes.

The purpose of this paper is to carry out such an investigation in the
asymptotically flat case.
In section 2, we review some of the basic thermodynamic and statistical
mechanical relationships that one would like to reproduce in the black
hole situation, highlighting the central role of the thermodynamic
potential $\Psi$ that corresponds to the statistical mechanical
partition function.
In section 3, we recall the Kerr-Bolt metric, and in section 4 we find,
unexpectedly, 
that its singularities can be removed only for certain, very special
values of the parameters.
In section 5 we find that asymptotic regularity in the presence of NUT
charge implies a closely related ``quantization'' of the parameters
of any thermodynamic ``reservoir'' that can be represented adequately by
boundary condtions on the gravitational path integral.
In section 6, we calculate the action-integral and the resulting
expressions for energy and angular momentum of the Kerr-Bolt spacetimes,
pointing out some unresolved ambiguities entering into the calculation.
In section 7, we discuss the status of the first law of 
thermodynamics 
with respect to 
these spacetimes, especally in light of the
parameter quantization found in sections 4 and 5.

\section{Review of thermodynamics: everything from $\log{Z}$}

First let us recall some thermodynamic relationships from flat spacetime.
The expression 
\begin{equation}
   \hat\rho = {1\over Z} \exp \braces{-\beta \hat{H} + \Omtilde \hat{J}} \ ,
\end{equation}
where  $\Omtilde=\beta\Omega$,
describes the Gibbs equilibrium state of a system in contact with a
environment/reservoir of temperature $T=1/\beta$ 
that rotates with angular velocity $\Omega$
around a specified axis. 
Here, $\hat{H}$ is the Hamiltonian operator
and $\hat{J}$ is the angular momentum operator, or rather its component
about the specified axis.
In addition, of course, the partition function $Z$ is given by
\begin{equation}
    Z = \trace \exp\braces{-\beta \hat{H} + \Omtilde \hat{J}} \ .
\end{equation}
It follows that, under variations of  $\beta$ and $\Omtilde$, one has
\begin{equation}
   d \log Z = - \ave{\hat{H}} \, d\beta + \ave{\hat{J}} \, d\Omtilde \ .
\end{equation}
Defining 
the {\it thermodynamic mass} $M=$ $\ave{\hat{H}}$
and 
the {\it thermodynamic angular momentum} $J$ $=$ $\ave{\hat J}$, 
and setting $\Psi=\log Z^{-1}$, 
we can write this as a relation among one-forms\footnote%
{In the following equations we denote the gradient operator on the
manifold of equilibria by the symbol $\partial$.  We reserve the symbol
`d' for infinitesimal variations (or for its role in expressing
integration).}
\begin{equation}
   \partial\Psi = M \, \partial\beta - J \, \partial \Omtilde \ ,
   \label{th1}
\end{equation}
or equivalently,
\begin{equation}
   M = \partialderiv{\Psi}{\beta}{\Omtilde} 
   \ ,  \ 
   -J = \partialderiv{\Psi}{\Omtilde}{\beta} \ ,
\end{equation}
For the entropy, we have
 $
   S = \trace \hat\rho \log {{\hat\rho}\,}^{-1} \ 
    \ideq \ave{\,\log {\hat\rho\,}^{-1}\,} \ 
     = \ave{\,\log Z + \beta \hat{H} - \Omtilde \hat{J}\,}
 $
or equivalently
\begin{equation}
  S = \beta M - \Omtilde J - \Psi \ .  
  \label{th2}
\end{equation}
This states that $S$ is the Legendre transform of $\Psi$ with respect to
$\beta$ and $\Omtilde$~:
\begin{equation}
   S = \beta   \partialderiv{\Psi}{\beta}{\Omtilde} 
    + \Omtilde \partialderiv{\Psi}{\Omtilde}{\beta}
    - \Psi \ .
 \label{ths1}
\end{equation}
Also,
\begin{equation}
\partial\Psi=M\partial \beta-J\partial(\beta\Omega)
            =(M -\Omega J) \partial\beta - \beta J \partial\Omega \ ,
\end{equation}
whence
\begin{equation}
   \partialderiv{\Psi}{\beta}{\Omega} = M-\Omega{J} \ ,
\end{equation}
whence
\begin{equation}
 S = \beta M - \Omtilde J - \Psi
   = \beta (M-\Omega{J}) - \Psi
   = \beta \partialderiv{\Psi}{\beta}{\Omega} - \Psi  \ ,
  \label{phantom}
\end{equation}
an alternative expression for the entropy that is useful in
understanding the origin of the area law \cite{sumati}.
Taking the gradient of (\ref{th2}), we find immediately
a final well known equation, 
\begin{equation}
   \partial S = \beta \partial M - \Omtilde \partial J \ .
           \label{th3}
\end{equation}

Although the simple derivations rehearsed
above take as their point of departure 
the Gibbs state (or in other words the canonical ensemble), 
the resulting relations among the thermodynamic variables 
$\Psi,S,M,J,\beta,\Omtilde$ (or $\Omega$),
can be valid more generally, 
and we will assume that this is the case,
specifically, for 
the relations
(\ref{th1}), (\ref{th2}), (\ref{ths1}), (\ref{phantom}) and (\ref{th3}).
(Remember in this connection that a canonical ensemble does not adequately
describe systems in which two
different thermodynamic phases are present simultaneously, nor can it be taken
literally for systems containing black holes.)

Now in a path integral calculation of the partition function, one
is formally summing over configurations which are periodic under a shift
of $i\beta$ in time and $i\Omtilde$ in angle.  One might 
perhaps
accomplish this
by an analytic continuation in both time and angle, but often it is
more convenient to continue in time alone, while keeping the angle real.
However, one must then do a subsidiary continuation in $\Omtilde$,
and the final result can be expressed as follows.
Let $\tau=it$ be the Euclidean time,
let $I_E=I_E(\gamma)$ 
be the Wick-rotated action of the path $\gamma$,
and let
$Z_E(\Delta\tau,\Delta\varphi)$ be the 
 partition
function, defined as the value of the path integral
with integrand $\exp\braces{-I_E(\gamma)}$, 
taken over all 
periodic
paths $\gamma$
whose period is given by
the Killing vector 
$\xi=\Delta\tau\,\ptl/\ptl\tau + \Delta\varphi\,\ptl/\ptl\varphi$.
The thermodynamic partition function $Z$ that figures in 
the general formulas given above
is then an analytic continuation of $Z_E$ to imaginary $\Delta\varphi$: 
\begin{equation}
    Z=Z_E(\beta,-i\Omtilde)  \ .         
    \label{Ztrue}
\end{equation}

Let us turn now to thermodynamic systems for which gravity is important.
For isolated gravitating systems, one customarily omits any explicit
environment, idealizing it instead as the region at infinity in an
asymptotically flat spacetime.  In this idealization the temperature,
angular velocity, etc. will be encoded in the asymptotic metric,
and
the
partition function 
$Z_E(\beta,\Omtilde)$ 
will be given by a
gravitational path integral in which the metric is required to satisfy
boundary conditions corresponding to the desired values of the
``intensive'' parameters $\beta$ and $\Omtilde$.  
We will return to the
correct definition of these boundary conditions in a later
section.\footnote
{The path integral used in the gravitational case is basically just
 copied from the non-relativistic case, even though the analytic
 continuation it presupposes has never really been understood, as far as
 we know.}

Once the boundary conditions are in place, the gravitational path
integral can be approximated using the saddle point method.  To
zeroth order the saddle point approximation is just the value of the
integrand at the saddle point, which in this case means that we
approximate $Z$ as $\exp(-I)$ where $I$ is the value of the Wick
rotated action at the saddle point or ``instanton'' solution,
analytically continued in $\Omtilde$ as described above.
Strictly speaking, one should include the ``one loop terms'' in the
saddle point approximation, since they are actually comparable in
magnitude to the contribution from $I$.  However we will follow the usual
practice of ignoring them, or rather of attributing their contribution
to ambient thermal radiation, and attributing the contribution from
$I$ to the horizon {\it per se}.  
We thus approximate, for the isolated black hole, 
\begin{equation}
   \Psi \approx I \ .
     \label{approx}
\end{equation}
With this approximation, 
our basic thermodynamic formulas become
\begin{equation}
  \partial I = M \partial\beta - J\partial\Omtilde \ ,
\end{equation}
and
\begin{equation}
   S = \beta M - \Omtilde J  - I
     =   \beta    \partialderiv{I}{\beta}{\Omtilde}
       + \Omtilde \partialderiv{I}{\Omtilde}{\beta} - I
    = \beta \partialderiv{I}{\beta}{\Omega} - I \ .
    \label{GDrel}
\end{equation}

\section{The Kerr-Bolt line element before identifications}

The spatio-temporal line element from which we will build our instantons
is given by 
\begin{equation}
  ds^2 = 
  -\frac{\Delta_{L}}{\rho_{L}^{2}}[dt+(2n\cos \theta -a\sin ^{2}\theta )d\varphi ]^{2}
  +\frac{\sin^{2}\theta }{\rho_{L}^{2}} [adt-(r^{2}+n^{2}+a^{2})d\varphi ]^{2}
  +\frac{\rho_{L}^{2}dr^{2}}{\Delta_{L}}
  +{\rho_{L}^{2}d\theta ^{2}} \ ,
\end{equation}
where
$ \Delta_L = r^{2} - 2mr + a^2 -n^2$ 
and
$ \rho_{L}^{2} = r^{2} + (n + a\cos\theta)^{2}$.
The corresponding line element of Euclidean signature is obtained by 
analytically continuing the $t$ coordinate, the NUT charge $n$ 
and
the rotation parameter $a$ to imaginary values:
\begin{equation}
 ds^{2}= \frac{\Delta_{E}(r)}{\rho _{E}^{2}}
        [dt-(2n\cos \theta -a\sin^{2}\theta)d\varphi]^{2}
     + \frac{\sin ^{2}\theta }{\rho _{E}^{2}}
        [adt-(r^{2}-n^{2}-a^{2})d\varphi]^{2}
    + \rho _{E}^{2}(\frac{dr^{2}}{\Delta _{E}(r)}+d\theta ^{2}) \ ,
   \label{line-elt}
\end{equation}
where
\begin{equation}
   \rho _{E}^{2} = r^{2}-(n+a\cos \theta )^{2} \ ,
\end{equation}
\begin{equation}
   \Delta _{E}(r) = r^{2}-2mr-a^{2}+n^{2} \ .
\end{equation} 
These metrics are actually special cases of a more general family of
metrics with cosmological constant, which we give for completeness in an
appendix.  However, since we will only study the asymptotically flat
setting, we will only need the metrics with zero cosmological constant.

We also remark here, that
(\ref{line-elt}) 
can be expressed more felicitously
in terms of the spinor coordinates of 
(\ref{zetacoords}) below,
if one takes advantage of invariantly defined one-forms such as 
$\zeta^\dagger\partial\zeta$
\cite{spinorref}.
Using that technique, one should
be able to carry out the removal of the singularities more perfectly
than we do below, 
but in this paper we will work with the more familiar polar
coordinates $r$, $\theta$,  $\varphi$, and $t$.

The above line element depends on three parameters,
$n$, $m$ and $a$.  
Equivalently, 
we can employ instead of the ``mass parameter'' $m$ 
the ``horizon radius'' $r_0=m+\sqrt{m^2+a^2-n^2}$,
the outermost root of $\Delta_E(r)=0$, 
which (for Euclidean signature) is the lower limit of the allowed range
of variation of $r$.
In terms of $r_0$ we have
\begin{equation}
   m = \frac{r_{0}^{2}-a^{2}+n^{2}}{2r_{0}} \ .
    \label{masss}
\end{equation}
Our parameters are then $n$, $r_0$ and $a$.  Notice that one parameter
(say $n$)
can be identified with an overall scale, whence the distinct ``shapes''
are determined by only two parameters.

For future reference we define also the ``angular velocity of the horizon''
\begin{equation}
  \Omega_H = 
  \left. -\frac{g_{t\varphi}}{g_{\varphi\varphi}}\right|_{r=r_{0}} =
  \frac{a}{r_{0}^{2}-n^{2}-a^{2}} \ ,
    \label{OmH}
\end{equation}
and the ``inverse horizon temperature'',
\begin{equation}
  \beta_H = \frac{4\pi(r_{0}^{2}-n^{2}-a^{2})r_{0}}{r_{0}^{2}-n^{2}+a^{2}} \ ,
    \label{betaH}
\end{equation}
related to the surface gravity
\begin{equation}
  \kappa =\frac{1}{2(r_{0}^{2}-n^{2}-a^{2})}\left. \frac{d\Delta_E}{dr}\right|
  _{r=r_{0}}=\frac{r_{0}-m}{(r_{0}^{2}-n^{2}-a^{2})}=\frac{%
  r_{0}^{2}-n^{2}+a^{2}}{2(r_{0}^{2}-n^{2}-a^{2})r_{0}} \ ,  \label{TNKdSsurfgra}
\end{equation}
by
$\beta_H/2\pi=1/\kappa$.


All the instantons we will consider will be derived from (\ref{line-elt})
by identifications.  As we will see, these identifications will not only
twist the imaginary-time direction near infinity, but also
modify even the topology of the 
``spatial''
two-sphere at infinity, somewhat in the
manner of the so called ``ALE instantons''.  That is, the topology at
infinity will (unfortunately) not necessarily be that of a circle
bundle over $S^2$.

\section{The non-singular instantons}

As written, our line element is singular along the polar axes
$\theta=0,\pi$ and at the ``horizon'' $r=r_0$.  
The ``string'' singularities at the poles 
have,
as is well known, 
a more complicated structure than
that belonging to
the familiar spherical coordinates for $S^2$,
and they cannot be removed
merely by the imposition of periodicity in azimuth $\varphi$.  
Rather two separate identifications are required to remove the two
``strings''. 
Moreover,  one needs a further
identification to remove the singularity at the horizon, 
and all of the identifications must be compatible.
This compatibility condition is crucial, but it seems to have been overlooked
in the literature on these metrics.  
To analyze it, we may refer everything to a lattice in the time-azimuth plane.

\subsection{Lattice analysis}

That  $\partial/\partial{t}$ 
and $\partial/\partial{\varphi}$ 
are
both Killing vectors makes it possible to quotient our metric by any
linear combination of the two (with constant coefficients).  More
generally, we can quotient by any {\it lattice} $\L$ of vectors which is
closed under vector sum and difference.  The resulting spacetime is
entirely determined by $\L$.  Let us consider how $\L$ must be chosen in
order to remove the two types of singularity we have to deal with.

To take advantage of the scale invariance of our family of metrics, we
will introduce in place of $t$ the dimensionless coordinate 
\begin{equation}
   \psi = t / 2n \ ,
\end{equation}
and the corresponding Killing vector
\begin{equation}
     \partial / \partial\psi = 2n \; \partial / \partial{t} \ .
\end{equation}

Now look at the neighborhood of the north polar axis, $\theta=0$.  
At the pole, the line-element acquires a degenerate direction.  
In order to compensate, 
we must quotient by a vector $\eta_+$ 
that is parallel to this degenerate direction 
and whose length is chosen so that
the quotient metric will not exhibit a conical singularity.
That is, the length of $\eta_+$ at a proper distance $\eps$ 
from the
north pole, must be $2\pi\eps$ to first order in $\eps$.  
Some algebra shows that the required vector is 
\begin{equation}
   \eta_{+} = 2\pi \, (\partial /\partial \psi + \partial /\partial \varphi) \ .
    \label{etaplus}
\end{equation}
This therefore is one of the vectors of $\L$, but we can say more.  It
is also necessary that no ``submultiple'' of $\eta_{+}$ belong to $\L$;
otherwise we would really be quotienting by a smaller vector.
That is, $\eta_{+}$ must be a ``minimal'' element of the lattice.
(By minimal we mean that $\lambda\eta_{+}\notin\L$ for
$0<\lambda<1$.)
The same analysis at the south pole furnishes a second minimal lattice
vector, 
\begin{equation}
   \eta_{-} = 2\pi (\partial /\partial \psi - \partial /\partial \varphi) \ .
    \label{etaminus}
\end{equation}

The singularity at $r=r_0$ can be analyzed similarly.  
Although the algebra is a bit messier,
it proceeds along the same lines and reveals
a third minimal lattice vector,
\begin{equation}
  \xi = \beta_H (\partial /\partial t + \Omega_H \text{ }\partial /\partial \varphi )
   \ ,
    \label{xidef}
\end{equation}
where $\beta_H$ and $\Omega_H$ were defined in
(\ref{betaH}) and (\ref{OmH}).

Finally, there are the special points where the polar axes meet the
horizon.  We believe that no extra conditions on $\L$ are
imposed by regularity at these special points.

What does impose a crucial extra condition, however, is the requirement
that our quotient spacetime be a manifold.  If the lattice $\L$ were
dense in the $\varphi$-$t$-plane, this obviously would not be the case,
as the quotient $\Reals^2/\L$ would be pathologically non-Hausdorff.  
In order to preclude this, one must arrange that three vectors 
$\eta_\pm$ and $\xi$ be {\it commensurate}.
(In other words, $\xi$ must be a linear combination of
$\eta_\pm$ with rational coefficients.)
Bringing in the minimality constraints as well (see appendix 2), 
one can see that the complete condition is
\begin{equation}
   p\xi = q_{+} \eta_{+} +  q_{-} \eta _{-} \ ,
   \label{lattice}
\end{equation}%
where $p$, $q_{+}$ and $q_{-}$ are three relatively prime integers:
$(p,q_{+})=(p,q_{-})=(q_{+},q_{-})=1$.  

Notice that these conditions have the exceptional solution, $p=q_+=1$,
$q_-=0$.  For this solution, $\xi$ coincides with $\eta_+$, but in all
other cases (except for the mirror image case $\xi=\eta_-$) all three
vectors point in distinct directions.  This exceptional solution will also
be exceptional in its thermodynamic properties.

{}From (\ref{lattice}), we find
\begin{equation}
   \beta_H = 4 \pi n\ {q_{+} + q_{-} \over p} \ ,
    \label{betaHc}
\end{equation}
\begin{equation}
   \Omega_H = \frac{1}{2n} {q_+ - q_- \over q_+ + q_-} \ ,
   \label{Om}
\end{equation}
and therefore
\begin{equation}
   \Omtilde_H = 2 \pi {q_{+} - q_{-} \over p} \ .
     \label{OtH}
\end{equation}

We emphasize that the parameters $p$, $q_\pm$ are integers and therefore
not continuously variable.
For this reason 
--- and in sharp contrast to instantons without NUT charge --- 
the solutions considered here fall into disconnected families
between which no continuous transition is possible.\footnote
{This must be related to the findings of \cite{holzegel} in the Lorentzian setting.}

\begin{figure}
\centering                        
\begin{minipage}[c]{.55\textwidth}
     \centering
     \includegraphics[width=\textwidth]{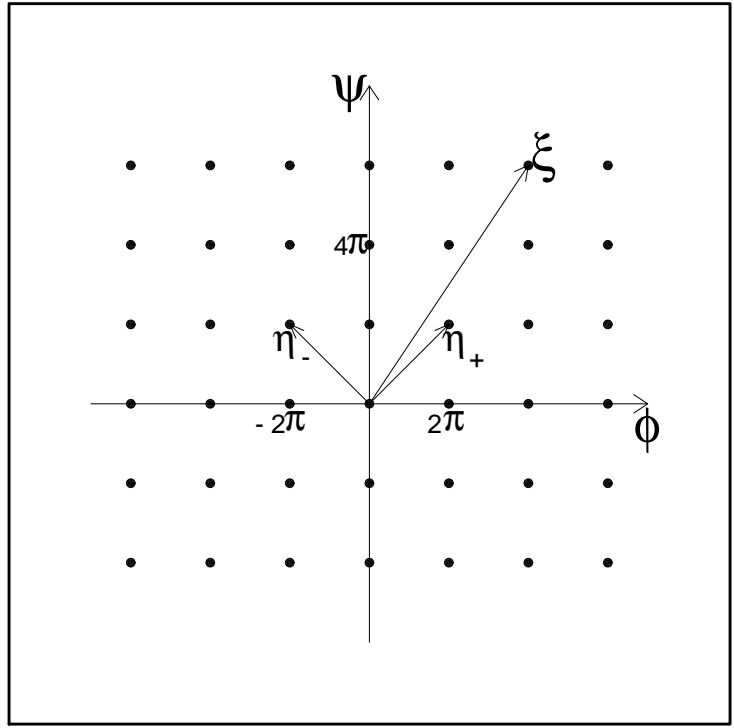}
\end{minipage}
\caption{{\it 
  The vectors $\eta _{+}$, $\eta _{-}$, $\xi$, and the lattice $\L$
  for the case $p=2$, $q_{+}=5$, $q_{-}=1$}}
\label{latticefig}
\end{figure}

\subsection{What is the topology near infinity?}

The identifications induced by the vectors $\xi$, $\eta_\pm$, produce a
Riemannian manifold that, near infinity, has (for fixed radius) the
topology of a lens space, or more particularly of the lens 
space $L(p; q_{+},q_{-})$.

We review some lens space lore in the second appendix.
By the expression `$L(p;q_1,q_2)$'  
we denote the quotient of 
the unit sphere $S^3\subseteq\Complexes^2$ 
by the action of the matrix, 
${\rm diag}(\1^{q_1/p},\1^{q_2/p})$, 
where $\1^y\ideq e^{2\pi iy}$
and the $q_i$ can be construed as
belonging to $\Integers_p$, the integers modulo $p$.
In fact the lens space
$L(p;q_1,q_2)$
depends only on the ratio $q=q_1/q_2$ 
--- taken in $\Integers_p$ --- 
and for
this reason it is commonly denoted simply by $L(p,q)$.
(In the exceptional case, $p=q_{+}=1$, $q_{-}=0$, 
 we would have $q=1/0=0/0$, since $\Z_1$ is itself zero, 
 and $1=0$ therein.  If we make the obvious convention  
 that $0/0=0$ in $\Z_1$, then our notation tells us that
 $L(1;1,0)=L(1,0)$,
 which of course is just $S^3$, since one is quotienting by the
 identity.) 
To grasp why $L(p; q_{+},q_{-})$
is the result of quotienting the
$\psi$-$\theta$-$\varphi$ coordinates by our lattice $\L$,
note first that 
an initial
quotienting by $\eta_{+}$ and $\eta_{-}$ 
produces
simply the 3-sphere, $S^3$, as explained, for example in \cite{KKmonopole}.
(The integral curves of the vector field $\eta_{+}+\eta_{-}$ then 
constitute the Hopf fibration of $S^3$ over $S^2$.)
Translation through 
the vector $\xi = (q_{+}/p) \eta_{+} +  (q_{-}/p) \eta_{-}$
can then be seen to be equivalent to multiplication by 
${\rm diag}(\1^{q_{+}/p},\1^{q_{-}/p})$
when $S^3$ is represented in terms of complex coordinates as above.
(These relationships appear with  particular clarity when expressed in
terms of the spinor coordinates alluded to earlier.)

For us, the important point is that different choices of the discrete
parameters $p$, $q_\pm$ yield in general different topologies near
infinity, because the lens spaces $L(p,q)$ are in general not homeomorphic.
(This makes it obvious that our collection of instantons
could not have formed a single continuous family.)
But that is not all.  
Except when $q=1$, $L(p,q)$ is not even a circle bundle over $S^2$.
It follows that near infinity the corresponding instantons do
not even possess the ``spatial'' topology of an $\Reals^3$; 
that is, they are not asymptotically flat, even in the topological sense.
Only for $q=q_{+}=q_{-}=1$,
meaning $\xi={1\over p}(\eta_{+} + \eta_{-})$,
do we have a full 2-sphere at infinity.   
(This solution would look near infinity like a monopole of charge $p$
if interpreted 
as a spatial hypersurface in a Kaluza-Klein spacetime.)
%
%
For all other $q$,
the fraction of $4\pi$ that remains turns out be
 $2/(q_{+}+q_{-})$,
independently of the value of $p$.

\REMARK  We have seen that not all values of $\beta$ and $\Omtilde$ are
accessible.  Correspondingly, where $I_E$ {\it is} defined, it fails to
be a single-valued function of $\beta$ and $\Omtilde$, because different
$L(p,q_+,q_-)$ can belong to the same $\Omtilde$ (cf. (\ref{IEuc1}) below).

\subsection{Parameter counting}

We have found in the last two subsections that the exclusion of
singularities rules out a large fraction of our original family of
(local) solutions (\ref{line-elt}).  How many free parameters remain?
Originally, we began with three continuous parameters, 
$n$, $r_0$ and $a$,
but our lattice
analysis revealed that a more appropriate starting point would have been
the three {\it integer} parameters  $p$, $q_\pm$.  
Once these are chosen,
two combinations of $n$, $r_0$ and $a$ are determined, 
as we have seen,
to wit: 
$\beta_H/n$ and $\Omtilde_H$.
Thus only a single continuous parameter remains.

The one exception occurs when $\xi=\eta_{+}$ (or $\eta_{-}$).
In that case, it turns out that (\ref{OmH}) and (\ref{betaH}) are not
independent conditions on $n$, $r_0$ and $a$, and there survives a
2-dimensional manifold of instantons, which can be parameterized by, for
example, $\beta_H$ and $r_0$.

In no case, though, can the parameters 
$\beta_H$ and $\Omega_H$ vary independently, since their product is
fixed to $2\pi(q_{+}-q_{-})/p$ by equation (\ref{OtH}).
For any other combination of horizon temperature and angular velocity,
no instanton is available to contribute to the thermodynamic partition
function $Z$ in the saddle point approximation.  
Although this suggests a pathological variation of $Z$ with
thermodynamic temperature and/or angular velocity, we will see next that
this would probably be an incorrect way to read the situation.  
The correct conclusion seems to be rather that
the thermodynamic ``reservoirs'' corresponding to the ``forbidden
points'' fail to exist altogether.

\section{Boundary conditions at infinity for a rotating, heated
  environment with NUT charge} 

As reviewed in section 2, a partition function $Z(\beta,\Omtilde)$
is meant to describe a system immersed in an environment which is
rotating with the angular velocity $\Omega=\Omtilde/\beta$.  
Whether such a picture really makes sense in a spacetime of NUT type can
certainly be doubted, because the string singularities associated with
the NUT charge (respectively, the closed causal curves that result if one makes
identifications to remove the strings) might be taken to render the
spacetime itself unphysical.
If this were the case, then 
the Wick rotation leading to the
Euclidean-signature path integral invoked earlier 
(whose justification is obscure in the best of cases, when gravity is involved) 
would also be meaningless, 
even if the instanton itself were free of singularities.
Despite such doubts, however, 
one can still hope that something of the usual
thermodynamic schema will continue to make sense in the presence
of NUT charge. 

Proceeding in this vein, let us ask what boundary conditions at infinity
could represent the type of reservoir we want.  
In an asymptotically flat spacetime (no NUT charge), 
the answer is clear-cut and derives from the
flat-space identifications described earlier.  
But since the line element we are working with is also flat 
near infinity --- at least locally --- we can hope to arrive at 
a unique set of boundary conditions for it as well
by analogy with the asymptotically flat case.  

For large radii, our line element (\ref{line-elt}) takes the form
\begin{equation}
 ds^2 = (dt - 2n\cos\theta d\varphi)^2
         + r^2 (d\theta^{2} + \sin^2\theta d\varphi^2) \ .
\end{equation}
The singularities at $\theta=0,\pi$  
are essentially the same as before,
and the same analysis goes through, 
telling us that 
in order to render the metric nonsingular,  
we must divide out by 
the same two vectors $\eta_+$ and $\eta_-$ 
that we found earlier.  
(With zero NUT charge, we would at this stage simply have divided out by
 $2\pi\partial/\partial\varphi$, 
 i.e. by nothing at all had our coordinates been Cartesian.)

This having been done, we can consider what further identifications will
express the thermodynamic boundary conditions we want.  
We saw earlier that in the absence of curvature
(i.e. in Wick rotated Minkowski spacetime)
one obtains $Z_E$ by identifying 
with respect to the vector\footnote%
{In section 2 we wrote `$\tau$' for `$t$' to avoid confusion between
 Lorentzian and Euclidean time.}
\begin{equation}
   \xi = 
   \beta \; \partial / \partial t + \Omtilde \; \partial / \partial\varphi \ ,
    \label{xi-def}
\end{equation}
which says that the manifold is to be made periodic with respect to
simultaneous shifts of the Euclidean time by $\beta$ 
and of angle by $\Omtilde$.
Here, we will simply {\it assume} that the same expression applies in
our situation.  
In respect of the first term, 
this assumption would seem to be on safe ground, 
since the Killing vector $\partial/\partial{t}$ is singled out as
the unique one whose norm goes to unity at infinity.  
We have no equally compelling argument for the second term, 
but we will assume it to be correct nonetheless.

The above equation may be taken as the {\it definition} of the
thermodynamic parameters $\beta$ and $\Omtilde$.  Physically the latter
represent ``intensive'' variables conjugate respectively to the energy
and the angular momentum, and they would of course be interpreted as
specifying the temperature and 
(after the analytic continuation (\ref{Ztrue})) 
the angular velocity of the reservoir.
Given this interpretation,
it is the requirement of global non-singularity that makes the
connection between the metric parameters $\beta_H$ and $\Omtilde_H$
defined at the horizon,  
and the thermodynamic parameters $\beta$ and $\Omtilde$ defined at
infinity (because it forces the vector $\xi$ to be the same
everywhere).\footnote
{This is actually an oversimplification, because all we really know is
 that the lattice of identifications $\L$ must be constant, and
 this leaves open the possibility that $\xi_H$ and $\xi_\infty$ could be
 different minimal vectors in $\L$.  In this way, one and the same
 instanton can apparently contribute to many different values of the
 partition function.  In this connection,  we remark also that 
 the definition of the Euclidean $\Omtilde$ suggests that, unlike its
 Lorentzian counterpart, it is only defined modulo $2\pi$.}

With the additional identification with respect to $\xi$ we are now
back in exactly the same situation as before, characterized by a lattice
$\L=\span\braces{\xi,\eta_+,\eta_-}$.  
The commensurability conditions are then also as before, 
and they lead to the analogous conclusion that the
manifold of possible equilibria is not connected:
we do not have a
reservoir for every possible choice of temperature and rotation rate!
In particular, we learn from  (\ref{xi-def}), (\ref{lattice}), and
(\ref{etaplus})-(\ref{etaminus})
that 
$\Omtilde=\beta\Omega$ cannot vary continuously at all;
it is fixed at the value 
\begin{equation}
   \Omtilde = 2\pi (q_{+} - q_{-}) / p \ .
\end{equation}
Obviously, this will make it difficult to define a thermodynamic mass,
angular momentum and entropy as we have done above, 
since the partial derivatives occurring in the definitions cannot all be
taken without going outside the manifold of equilibria.

\REMARK  If we were willing to let $\eta_\pm$ vary, then we could
introduce a second free parameter.  However, the resulting singularities
at the poles (which would be conical singularities or worse) would
result in divergent expressions for the Euclidean action $I$.

Finally, it's worth asking what went wrong.  
Why, for example, is it possible to vary $\beta$ and $\Omtilde$
freely and independently for the Kerr instanton?
The answer, 
as intimated earlier, 
lies in the nature of the lattice $\L$.  
Without NUT charge,
the coordinate singularity of polar coordinates can be canceled
by a single identification with respect to the lattice vector
%
%
$2\pi \, \partial/\partial\varphi$,
leaving the vector $\xi$ that carries the thermodynamic information 
free to roam unmolested through the entire $t$-$\varphi$-plane,
because, in a lattice with only two\footnote
{One could equally well say that there are still three generating vectors, but
 they reduce effectively to two, since the identifications needed
 at the north and south poles are the same.  In this way of looking at it,
 turning off the NUT charge ``flattens'' these two lattice vectors so
 that they become proportional:  $\eta_{-}=-\eta_{+}$.} 
generating vectors, 
no incommensurability can arise.  
For all choices of $\xi$, 
the quotient topology is then simply $S^2\times S^1$.
But with NUT charge, 
the angular singularities of the line element are altered,
and two {\it independent} identifications $\eta_\pm$ are needed to cancel them.
The ``thermodynamic vector'' $\xi$ is then restricted to a discrete set
of values, 
each yielding one of the lens space topologies.

\section{The Euclidean action, ``conserved'' mass and angular momentum}

In order to apply the thermodynamic definitions of Section 2 to 
our family of instantons, we need to define the Euclidean action $I$. 
Once we have done so, we can also appeal to Noether's method to 
derive a second pair of definitions of mass $\M$ and angular momentum $\J$,
logically distinct from the ``thermodynamic'' quantities defined
earlier.\footnote
{We use ``fraktur'' letters to distinguish these quantities from 
 the corresponding ``thermodynamic'' quantities defined earlier.
 We will call the first set of quantities ``thermodynamic'' and the
 second set ``conserved'' 
 (despite the doubts about conservation expressed below).}
For consistency of the analysis, one would hope to obtain the same
values, no matter which definition one used, but we will see below that
things are not quite that simple when NUT charge is present.

In Minkowski spacetime, Noether's
theorem furnishes us with a set of quantities
derived directly from the action principle underlying the dynamics,
and conserved in the sense of taking the same value 
no matter which Cauchy surface they are evaluated on.
In a gravitational setting both the action and the conserved quantities
must be defined relative to an ``environment at infinity''.  
The corresponding boundary conditions are delicate however,
and people have devised several distinct versions of
the definitions for application to curved spacetimes. 
In addition to the counteterm method that we will employ here, 
we mention in particular the method of \cite{actionvar} and \cite{madh}.
In \cite{joohan} and \cite{gaboretal},
the earlier form \cite{actionvar} of this method 
was successfully applied to the Kaluza-Klein monopole,
a spacetime whose topology resembles that of 
the gravitational instantons we are studying in the present paper.  
It would be interesting to apply its later form \cite{madh} to some of
the solutions with NUT charge.

Unfortunately, the Noetherian method, when carried over to spacetimes
with NUT charge, has trouble rooted in the topological ``twist''
characterising such spacetimes.
Either a string singularity 
is present or else closed timelike curves.  
In the former case, one encounters divergent integrals, 
which have to be ``renormalized'' somehow;
in the latter case, the integrands are finite, but the surface over
which one wishes to integrate gets ``torn''.  
For present purposes, we need to set up mass and angular momentum
integrals for unphysical metrics of Euclidean
signature (instantons), but it turns out that the topological
obstruction to doing so is essentially the same as in the Lorentzian
case. 

In defining the gravitational Euclidean Action, 
one integrates over the entire spacetime, 
and the fact that 
the topology at infinity is that of a twisted circle bundle
presents no particular problem of principle.  
But the topology does present a problem when
one tries to define mass and angular momentum.  
The integrals defining these these quantities
ought to extend only over a {\it section} of the bundle, 
but in a spacetime with NUT charge, no such section exists.  
If one ignores this problem and integrates, 
over a surface with boundaries, 
then the result depends on which surface one chooses
because ``flux can leak through the gaps''.  
This means that the mass $\M$ and angular momentum $\J$ 
one ends up with are not really conserved in the sense
of being independent of integration surface.
 
In light of the above, it seems fair to say that the definitions of mass
and angular momentum in NUT-type spacetimes are inherently ambiguous
(mirroring, perhaps, the ambiguities of definition in the thermodynamic
 angular momentum).
In the following we will use the values that result from a seemingly
natural choice of integration surface, namely a surface of constant $t$.

In applying the counterterm method, 
we will temporarily allow for a nonzero
cosmological constant which we will then send to zero in order to obtain
values appropriate to the asymptotically flat case.
We refer the reader to the first appendix for
the notation and the explicit form of the metric. 
To compute $\M$ and $\J$, we consider
the four dimensional action that yields the Einstein equations with a
negative cosmological constant 
\begin{equation}
  I=\frac{1}{16\pi }\int_{{\cal M}}d^{4}x\sqrt{-g}\left( R-2\Lambda \right) -%
  \frac{1}{8\pi }\int_{{\cal \partial M}}d^{3}x\sqrt{h}K+I_{ct} \ ,
  \label{totaction}
\end{equation}%
where ${\cal \partial M}$ is the boundary for constant (large) $r$, and $%
\int_{{\cal \partial M}}d^{3}x$ indicates an integral over this boundary,
whose induced metric and extrinsic curvature are respectively $h_{\mu\nu}$
and $K_{\mu\nu}$ induced from the bulk spacetime metric $g_{\mu\nu}$. 
$I_{ct}$ is the counter-term action, calculated to cancel the divergences
from the first two terms, and we have set the gravitational constant
$G=1$. 
The associated boundary stress-energy tensor $T^{ab}$ is obtained by the
variation of the action with respect to the boundary metric 
$T^{ab}=\frac{\delta I}{\delta h_{ab}}$, 
the explicit form of which can be found in \cite{das, stressten}. 

If the boundary geometries admit an isometry 
generated by a Killing vector $\chi^{a}$, 
then it is straightforward to show that $T^{ab}\chi_{b}$ is
divergenceless, from which it follows --- at least locally ---
that 
there will be a ``conserved charge'' 
${\Q}$ between surfaces of constant $t$.
%
%
If $\partial/\partial{t}$ is itself the Killing vector, then 
${\Q}={\M}$, 
the conserved mass associated with the boundary surface $\Sigma$ 
at any given ``time''.  


For the Killing vector $\partial/\partial{t}$ we 
can obtain the conserved mass,
from the Kerr-Bolt-AdS metric near infinity. 
[The Kerr-Bolt-AdS metric can be obtained by analytic continuation of
the cosmological parameter in the metric (\ref{LKBdS}) or (\ref{EKBdS})
from postive to negative values.] 
For vanishing cosmological constant, this reduces (in the limit of
infinite radius) simply to the mass parameter $m$.
The other conserved charge,
associated with the axial Killing vector $\partial/\partial\varphi$, 
is the ``conserved angular momentum'',
$  {\J}=\frac{am}{\Xi^{2}}  $,
which vanishes when $a=0$, 
and reduces to $am$ 
in the limit of vanishing cosmological constant.

Finally, the action itself 
for the Kerr-Bolt metric 
is given by\footnote%
{The values (\ref{IE}), (\ref{Mcons}) and (\ref{Jcons}) can also be
 obtained directly in the asymptotically 
 flat setting, without introducing a nonzero cosmological constant.}
\begin{equation}
   I_E = { 4 \pi n m \over p} \ .
    \label{IE}
\end{equation}

\REMARK In arriving at (\ref{IE}), one needs to do a trival integral 
 over $\varphi$ and $t$.  We have used for the value of that integral 
 not $2\pi\beta$, as one might naively expect to obtain, but the area of 
 a ``unit cell'' of our lattice, namely $16\pi^2n/p$.

As explained above, we have computed ${\M}$ and ${\J}$
ignoring the fact that our surfaces of constant $t$ are both
``multivalued'' and singular at the poles.  The latter difficulty we
have agreed to ignore for now, 
but the multivaluedness can and should be
corrected for 
by introducing the factor $2/(q_{+}+q_{-})$ obtained 
in Section 4.2 
above. 
When we do so, we find 
\begin{equation}
 {\M} =  {2\over q_+ + q_-} m \ ,
     \label{Mcons}
\end{equation}
and
\begin{equation}
 {\J} =  {2\over q_+ + q_-} a m \ .         
    \label{Jcons}
\end{equation} 

In computing $I$,
there was no ambiguity of the sort 
that affected ${\M}$ and ${\J}$, 
but unfortunately, there is a different kind of ambiguity.
As explained earlier in connection with equation (\ref{Ztrue}), 
the action $I_E$ belongs in effect to a purely imaginary value of
the angular velocity.  
In order to obtain the
true thermodynamic potential $\Psi$,
we would need to continue  from $\Omtilde$ to $-i\Omtilde$,
but the desired analytic continuation is ambiguous because
$I_E(\beta,\Omtilde)$ is defined only for a 
discontinuous set of $\Omtilde$,
and where it is defined, it is a multi-valued function of $\beta$ and
$\Omtilde$ (as follows from (\ref{IEuc1}) below).
Clearly, this ambiguity will also affect the definition of the
true
thermodynamic $M$ and $J$.
Once again the topology is getting in the way of the thermodynamics!

For now, let us just record the value of the pre-continuation action
(\ref{IE}).
Expressed as a function of $p$, $q_\pm$, and $n$, it is
\begin{equation}
   I_E 
  = 
   4\pi n^{2} 
   \; 
   {p \; + \; 4 \qplus \qminus / p
      \over
    p (\qplus + \qminus) + \sqrt{4\qplus\qminus (p^2 - (\qplus - \qminus)^2)}}
    \ ,
     \label{IEuc}
\end{equation}
%
%
which can also be written simply in terms of 
the rationalized dimensionless variables 
$\betahat=\beta/4\pi n$, 
$\Omegahat=\Omtilde/2\pi$
as
\begin{equation}
  I_E
  =
  {4\pi n^2\over p}
  {1 + \betahat^2 - \Omegahat^2
   \over
  \betahat + \sqrt{(1-\Omegahat^2) (\betahat^2-\Omegahat^2)}} \ .
     \label{IEuc1}
\end{equation}
Notice that for our exceptional solution 
(where  $p=\betahat=\Omegahat=1$)
this reduces to 
\begin{equation}
   I_E = 4\pi n^2 = {\beta^2 \over 4 \pi} \ .
  \label{IEucEx}
\end{equation}

%

\section{Do the thermodynamic relationships hold?} 

Does the ``first law'' hold for rotating black holes with NUT charge?
If it is meant to include the normal relationship between horizon area
and entropy, namely
\begin{equation}
    S = A/4G = 2\pi A/\kappa \ ,                    
    \label{arealaw}
\end{equation}
(where  $\kappa=8\pi G$ is the rationalized gravitational constant),
then
it fails already in the non-rotating case, $a=0$ \cite{robb}.  
With nonzero $a$, 
it seems at first sight to fail even more abjectly, because one
cannot even define a consistent entropy on the basis of the formula
(\ref{th3}) together with the formulas for the ``conserved'' mass and angular
momentum given in 
Section 6.
However, the question is complicated by the unusual nature of the
manifold of equilibrium states, 
as it emerged in Section 4 above, 
which is not simply parameterized 
by the quantities $\Omega$ and $\beta$.  
Given this, it seems best to begin by reconsidering the
meaning of the first law in as general a setting as possible.

To that end, let $\E$ be the manifold of thermodynamic 
equilibrium states of some system for which the only relevant
``environmental'' parameters are 
the (inverse) temperature $\beta$ and the angular velocity $\Omega$.
(Notice that this does not imply that $\dim\E=2$.  The case $\dim\E>2$
is perfectly possible for a system undergoing a phase transition; and
$\dim\E=1$ also makes sense if $\beta$ and $\Omega$ fail to be
independent.) 
Normally, we would expect a 
number
of ``state functions'' 
to be defined on $\E$,
including 
the thermodynamic ``potentials'' $\Psi=\log{Z^{-1}}$ and $S$, 
the intensive quantities $\beta$ and $\Omtilde$, 
and the corresponding ``extensive'' quantities $M$ (or $E$) and $J$.  
Among these six scalar fields on $\E$ there subsist, normally, three
basic relationships recalled earlier 
as equations (\ref{th1}), (\ref{th2}) and (\ref{th3}):
\begin{equation}
   \partial\Psi = M \, \partial\beta - J \, \partial\Omtilde \ ,
      \label{th11}
\end{equation}
\begin{equation}
  S = \beta M - \Omtilde J - \Psi \ ,
      \label{th22}
\end{equation}
\begin{equation}
   \partial S = \beta \partial M - \Omtilde \partial J \ .         
      \label{th33}
\end{equation}
We may thus interpret the question whether the first law
holds as asking whether these three equations are verified.\footnote
{In order that the points of $\E$ represent genuine equilibria, it is
 necessary that (\ref{th33}) hold not only on tangent vectors within
 $\E$ but also for variations that lead off of $\E$; i.e. it must hold
 as an equation among 1-forms in the larger 
 state-manifold, 
 not just when
 pulled back to $\E$.  In addition, the equilibrium will be unstable
 unless certain subsidiary conditions of a quadratic nature are
 fulfilled.  (For discussion of some of these points see for example
 \cite{turningpoint} and \cite{winnipeg}.)  To our knowledge these
 secondary conditions have never been verified for the black hole case,
 not even for a geometry as simple as that of the Schwarzschild metric.
 Doing this would place black hole thermodynamics on an even firmer
 footing than it already rests on.}
Or we may
interpret it more broadly to require, in addition, the area law
(\ref{arealaw}). 

Of course these equations are not all independent, 
because (\ref{th11}) and (\ref{th33}) are equivalent in the presence of (\ref{th22}).
For example, suppose we
begin with a family of Euclidean-signature instantons parameterized by
$\beta$ and $\Omtilde$ and set $\Psi=I_E(\beta,-i\Omtilde)$.  
If we
then define $M$ and $J$ from (\ref{th11}) 
[assuming this is possible]
and $S$ from (\ref{th22}),
then (\ref{th33})
becomes a tautology; there is nothing to check except possibly (\ref{arealaw}).
On the other hand, if we know independently any of $S$, $M$ and $J$,
then there exist consistency conditions to be satisfied.  In addition,
equation (\ref{th11}) itself implies a certain consistency condition 
since it implies that $\Psi$ is (locally) a function of $\beta$ and
$\Omtilde$ alone, 
and this is not automatic 
when $\dim\E>2$ 
or when $\partial\beta$ and $\partial\Omtilde$ fail to span 
the cotangent space to $\E$. 

Now let us turn to the case of interest in this paper.  
As we have seen,  the possible ``reservoirs at infinity'', 
and therefore the possible equilibria, fall into families 
parameterized by the relatively prime integers $p$, $q_\pm$.
Let $\E(p,q_\pm)$ be the equilibrium 
submanifold 
belonging to given 
$p$, $q_\pm$, or rather the set of instantons belonging to them, since
we are working in the approximation (\ref{approx}).  
What is peculiar about our situation is that
$\Omtilde$ is {\it constant} on each submanifold  $\E(p,q_\pm)$, as we learned in
Section 4.
Its gradient
$\partial\Omtilde$ is thus zero and the basic equation (\ref{th11})
reduces simply to
\begin{equation}
   \partial\Psi = M \, \partial\beta \ ,
    \label{dpsi}
\end{equation}
while (\ref{th33}) also simplifies, to
\begin{equation}
   \partial(S + \Omtilde J) = \beta \partial M \ .
     \label{dcomb}
\end{equation}
Equation (\ref{th22}) is not affected as such,
although $J$ can no longer be read off from 
$\partial\Psi$ 
and must come from somewhere else,
assuming it is defined at all.
The conditions we have to consider in connection with the first law are
thus (\ref{dpsi}), (\ref{th22}), (\ref{dcomb}), (\ref{arealaw}).
Implicit in these
as well are the conditions of agreement between the thermodynamic
quantities $M$ and $J$ (when they are defined) and their
analogs ${\M}$ and ${\J}$ as given by 
(\ref{Mcons}) and (\ref{Jcons}):
\begin{equation}
    M = {\M} , \quad J = {\J} \ .
    \label{ConsConsis}
\end{equation}

For any given 
equilibrium family
$\E(p,q_\pm)$, we may thus pose a sequence of questions
to bring out the extent to which the above thermodynamic 
relationships 
are verified. 
Beginning with $\Psi$ as approximated --- modulo analytic continuation! --- by 
(\ref{IEuc}) or (\ref{IEucEx}), 
we can ask whether (\ref{dpsi}) holds\footnote
{That (\ref{dpsi}) holds says basically that $\Psi$ (and consequently $M$)
 is a function of $\beta$ alone.}
for some $M$.
If it does, we can take this to define the ``thermodynamic mass'',
and we can further solve (\ref{dcomb}) to give meaning to the combination,
$S + \Omtilde J$.  
But this is as far as we can go on the basis of $I$ alone,
since (\ref{dpsi}) obviously does not let us define a thermodynamic angular
momentum $J$. 
Consequently, we cannot define a separate $S$ either, and 
therefore we cannot 
give meaning to (\ref{th33}) or
check the area law for entropy.\footnote
{We could possibly define a thermodynamic angular momentum $J$ if we
 could somehow extend $\Omtilde$ off of $\E$ (or extend $\E$ itself),
 say by admitting conical
 singularities in the metric near infinity, but 
 that would likely
 introduce divergences into $I$.  Even if this failed, it might still be
 possible to bypass the definition of $J$ and define an entropy $S$
 directly, if we could carry over to the Kerr-NUT case the technique of
 \cite{sumati}, based on ``shrinking the metric along $\xi$''}

In order to proceed further, 
we can bring in our independent knowledge of mass and angular momentum
in the role of ``conserved quantities''.
Specifically, we can ask whether the thermodynamic mass
$M$ (assuming it was well-defined) agrees with the ``conserved mass''
${\M}$.  If this consistency check succeeds, we can 
go further and simply {\it define} $J$ to be ${\J}$.
Having done so, we can obtain $S$ from (\ref{th22}). 
[Eq (\ref{dcomb}) will be guaranteed to hold.]  
We can then ask whether (\ref{arealaw}) holds.

Finally we can proceed in the opposite direction,
starting from (\ref{th33}), with $M$ and $J$ now taken to be
${\M}$ and ${\J}$, respectively.  
When (\ref{th33}) is integrable
(a nontrivial condition only in the 2-dimensional subcase), 
we can deduce an entropy $S$ from it,
this being, of course, the most
venerable order in which to proceed.
And we can then obtain $\Psi$ from (\ref{th22}).
Having done so, we can check $\Psi$ for consistency with $I$,
and we can compare $S$ with the area.

An analysis along these lines
falls naturally into two subcases,
the generic case in which $\E(p,q_\pm)$ has only a single degree of
freedom, and the exceptional case of $\E(1,1,0)$, where two
independent parameters remain free.  
Let us begin with the latter, since it seems to
hold the greater interest.

The exceptional family of instantons is characterized by the equation 
$\xi=\eta_+$ (or equivalently $\xi=\eta_-$), corresponding to the
parameter values $p=1$, $\qplus=1$, $\qminus=0$.
For these values, 
we see from (\ref{OtH}) and (\ref{betaHc}) that
\begin{equation}
   \Omtilde=2\pi \ , \quad \beta = 4\pi n \ ,\quad\Omega = 1/2n \ ,
\end{equation}
and the solution of (\ref{OmH}) and (\ref{betaH}) is 
\begin{equation}
                 r_0 = n + a \ ,           
    \label{r0}
\end{equation}
(which solves both (\ref{OmH}) and (\ref{betaH}) simultaneously, leaving free
both of the 
parameters $n$ and $a$).  We thus obtain a two-parameter family of 
solutions.\footnote
{assuming nothing goes wrong where the horizon meets the polar axes}
Substituting (\ref{r0}) into (\ref{masss}) yields for the ``mass parameter'' $m$,
simply $m=n$.
According to (\ref{IEucEx}), the Euclidean action is
\begin{equation}
         I_E = 4 \pi n^2  \ ,
\end{equation}
and {\it  if we ignore the issue of analytic continuation in $\Omtilde$},
we can equate this to $\Psi$, following equation (\ref{approx}).
The ``conserved'' mass and angular momentum reduce for these solutions
to\footnote%
{Strangely, $\M$ does not take on the Taub-bolt value of 
 $\frac{5}{4}n$ when $a\to0$.}
\begin{equation}
         \M = 2n \ ,\quad \J = 2an \ .
\end{equation}
Finally the horizon area,
computed from $ds^2$ with $r=r_0$, $t$=constant, 
and $\varphi$ in the range $[0,4\pi/(\qplus+\qminus)]$
is\footnote
{We remind the reader of the ambiguities exposed above which affect the
 values of $\M$, $\J$ and $A$, especially the value of $\J$, which
 involves both the question of integration surface and the question of
 analytic continuation.  The ambiguity in $A$ is similar to that in $\M$
 but milder thanks to the null character of the horizon (in both the
 Lorentzian and Euclidean settings).}
\begin{equation}
   A = 16 \pi n a \ .      
   \label{Ax}
\end{equation}


Now to what extent are these values compatible with our thermodynamic
formulas?  To start with, the proportionality 
of $\partial\Psi$ to $\partial\beta$ 
(equation (\ref{dpsi})) is not a foregone conclusion, since our
solution manifold is 2-dimensional.  
However, this first test is clearly passed,
because both $\Psi$ and $\beta$ depend on $n$ alone.  
The thermodynamic mass $M$ is thus well-defined and given by 
\begin{equation}  
    \ptl \Psi = M \ptl \beta \qquad \implies \qquad M=2n  \ .
\end{equation}
As remarked above, we cannot deduce from $\Psi$ a thermodynamic
angular momentum $J$ or an entropy $S$, but we can conclude that
\begin{equation}
  \ptl(S+\Omtilde J) = \beta\ptl M = 4\pi n(2\ptl n) = 8\pi n \, \ptl n \ ,
\end{equation}
from which it follows that, up to a constant of integration which we may
set to zero,
\begin{equation} 
       S + \Omtilde J = 4\pi n^2 \ .
\end{equation}
The first law (\ref{th3}) then holds trivially, for any choice of $J$, if we
set
$S=4\pi n^2-\Omtilde J$.  
The first law is effectively a 
tautology 
if approached from this
direction. 

What is not tautological, though, is the agreement (or lack thereof)
between the ``thermodynamic'' and ``conserved'' masses.
Somewhat remarkably, they do agree: $M=\M=2n$.  
(Notice that this is twice the value of the parameter $m$.)
This means that 
(as a two-line computation corroborates)
we would have arrived at the same $S$ had we begun from
the first law (\ref{th33}) and plugged in $M=\M$, $J=\J$.
If we adopt these values, then we can also obtain $S$ from 
(\ref{th22}), with the result
\begin{equation} S = \beta M - \Omtilde J - \Psi \ ,          \end{equation}
\begin{equation}   = 4\pi n(2n - a - n) \ ,       \label{e61} \end{equation}
\begin{equation}   = 4\pi n (n - a) \ .                       \end{equation}
By way of comparison, the area given in (\ref{Ax}) is
$ A = 16 \pi n a $.  Clearly, $S\not=\fourth A$.
Rather their ratio is 
\begin{equation}
   {S \over A/4 }  = {n-a \over a}  \ ,   
    \label{Sratio}
\end{equation}
which varies from $-1$ to $\infty$ depending on the ratio $n/a$.

In summary, only three of our ``tests'' were nontrivial in this case:
that $\ptl\Psi\propto\ptl\beta$, 
that $M=\M$, 
and (if we wish to include it)
that $S=A/4G$.
The first two were passed, but not the third.

\REMARK That $S$  can be negative suggests that something is
 wrong.  If the long sought analytic continuation were to reverse the sign
 of $\Omtilde J$, as it normally does 
 (and if it had no further effect on $\Psi$!), then
 the second term in (\ref{e61}) would change sign and
 we'd have instead of (\ref{Sratio}),
\begin{equation}
            {S \over A/4 }  = {n+a \over a}   \ge 1 \ .
\end{equation}

Now let us turn to the thermodynamics of the generic solutions.
Here even fewer nontrivial tests remain, since there is a single free
parameter and almost all the relevant results follow from dimension
counting.   The only relationship that could go wrong (aside from
(\ref{arealaw}) whose failure is already familiar) would be equality of the
thermodynamic and conserved masses.  However, this agreement follows
very simply 
from the form of equations (\ref{IEuc1}), (\ref{Mcons}) and (\ref{betaHc}).
According to the first of these equations, $I_E$ takes the form
\begin{equation}
   \Psi = {4\pi\over p} f(\betahat,\Omegahat) n^2 \ ,
\end{equation}
which, except for the prefactor $4\pi/p$, is determined by the fact that
Action is a length squared.  From  (\ref{Mcons}) we have similarly
\begin{equation}
   \M = {2\over \qplus + \qminus}  f(\betahat,\Omegahat) n \ ,
\end{equation}
for the same function $f$,
as one sees from (\ref{IE}) and (\ref{Mcons}).
And for $\beta$ we have from  (\ref{betaHc}),
\begin{equation}
   \beta = {4\pi\over p} {q_{+} + q_{-} \over 1} n \ .
\end{equation}
These three equations imply immediately that
\begin{equation}   
   \ptl \Psi = \M \;  \ptl\beta \ ,
\end{equation}
independently of the detailed form of $f(\betahat,\Omegahat)$.  
We see in particular that the same equality would hold if we analytically
continued $f$ to $f(\betahat,-i\Omegahat)$.  

\begin{figure}[tbp]
\centering    
\begin{minipage}[c]{.45\textwidth}
         \centering
         \includegraphics[width=\textwidth]{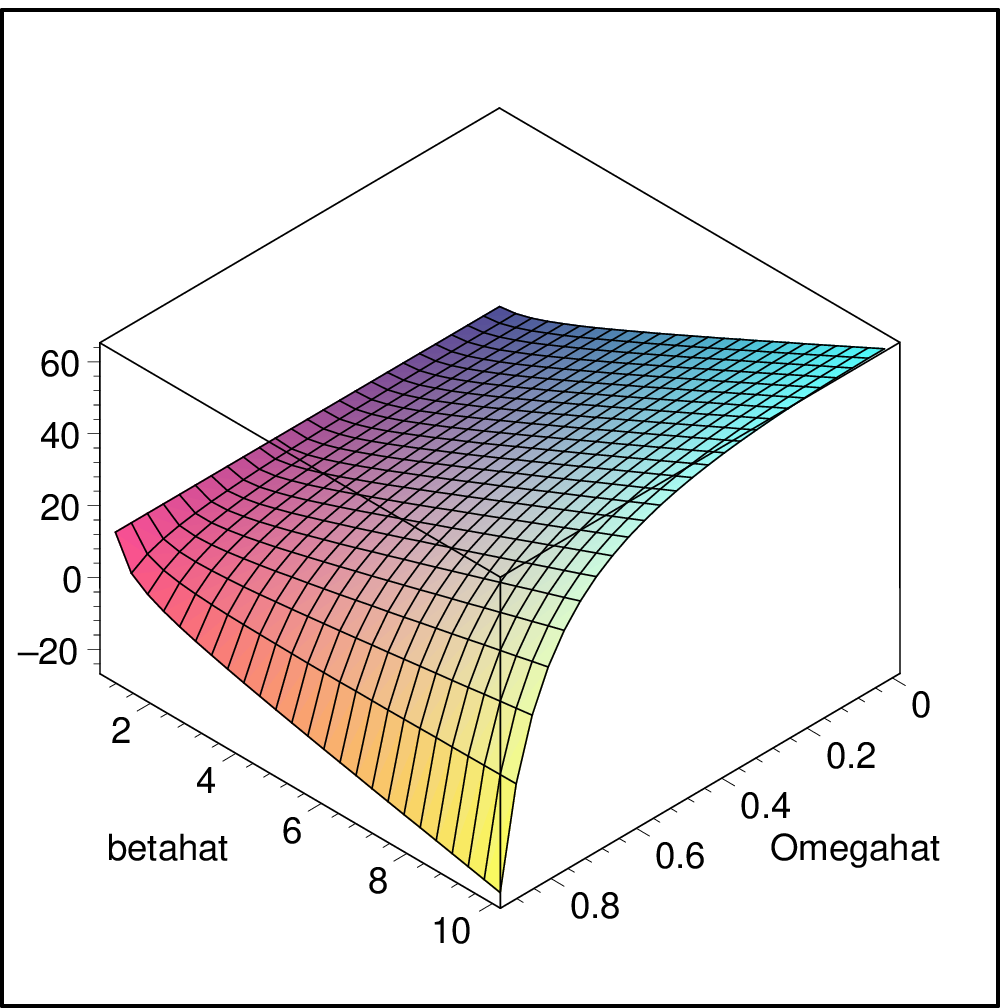}
         \caption{Entropy as function of $\betahat$ and $\Omegahat$ where
          we set $n=p=1$.}
         \label{Entropy}
\end{minipage}\begin{minipage}[c]{0.05\textwidth}
\end{minipage}%
\begin{minipage}[c]{.45\textwidth}
         \centering
         \includegraphics[width=\textwidth]{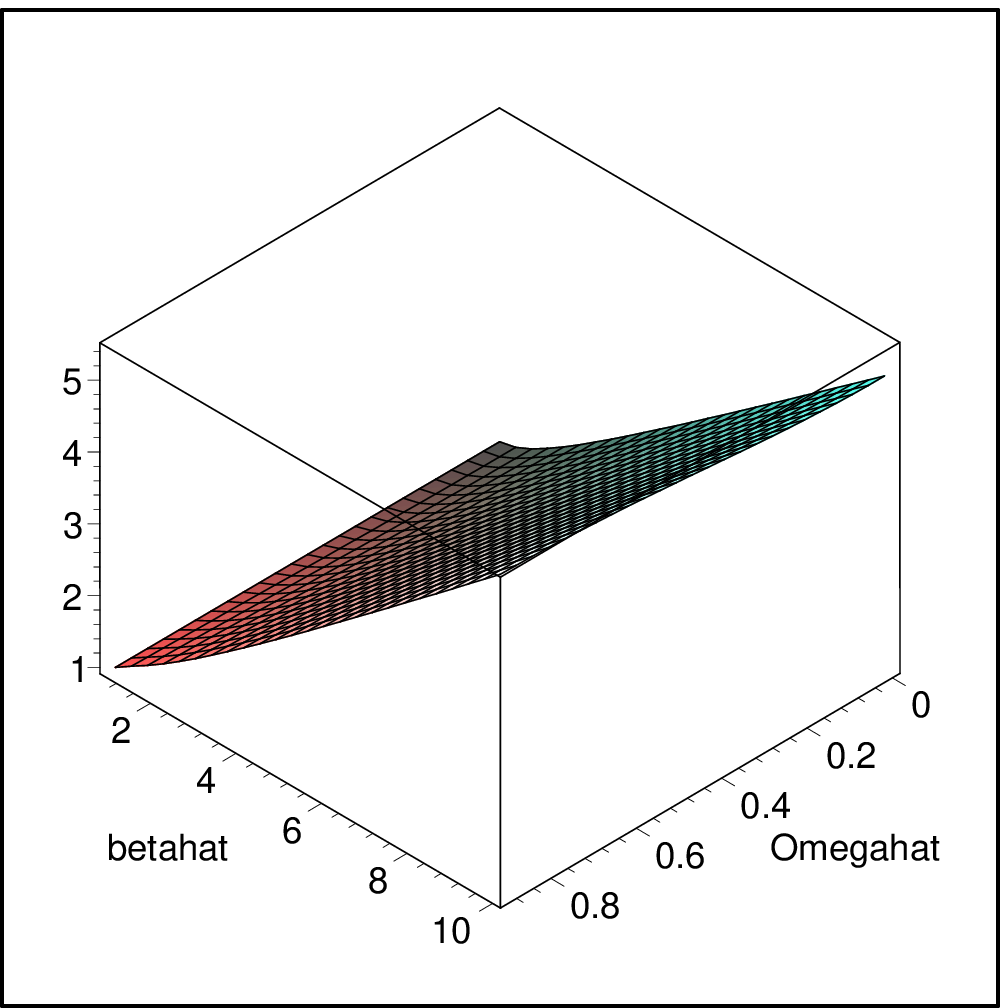}
         \caption{Entropy after analytic continuation as function of
         $\betahat$ and $\Omegahat$
          where we set $n=p=1$.}
         \label{Entropyaft}
\end{minipage}

\end{figure}

Following along as before, we can deduce
\begin{equation}
         S + \Omtilde J = {4\pi n^2 \over p}f(\betahat,\Omegahat) \ ,
\end{equation}
where the explicit form of the function $f$ is
\begin{equation}
  f(\betahat , \Omegahat)
  =
 {1+\betahat ^2-\Omegahat ^2
 \over 
 \betahat + \sqrt{(1-\Omegahat^2)(\betahat^2-\Omegahat^2)}} \ , 
\end{equation}
as can be read off from the equations referred to above.
If we identify $J$ with $\J$, 
this yields for the entropy
\begin{equation}
   S = 
   {4\pi n^2 f(\betahat,\Omegahat) \over p \betahat}
   \left(
     2\betahat - \sqrt {{\betahat ^2 - \Omegahat^2} \over 1-\Omegahat^2}
   \right) 
   \ .
\end{equation}
%
The entropy is
positive for $0 \leq \Omegahat \leq {\sqrt{3} \over 2}$ 
(for $\betahat \gg 1$)
while
for $\Omegahat > {\sqrt{3} \over 2}$ it takes positive as well as negative
values, depending on the value of $\betahat$ (figure (\ref{Entropy})).  
We note that one can show that\footnote
{Strictly speaking, this restriction only applies to $\beta_H$ and
 $\Omtilde_H$, not directly to the corresponding thermodynamic
 parameters as such. (See footnote \eight above.)}
$0 \leq \Omegahat^2 \leq 1 \leq \betahat$,
with $\Omegahat=1$ only if $\betahat=1$.
These inequalities can be expressed equivalently as 
$|q_{+} - q_{-}| \le p \le q_{+} + q_{-}$.
(Once again the analytic continuation in $\Omega$ could change the
 story.  If we naively substitute $-i\Omegahat$ for $\Omegahat$, the
 resulting expression for $S$ is manifestly positive.)
By way of comparison, the horizon area is given by\footnote 
{For $\Omegahat=\betahat=1$, corresponding to the exceptional
 family of instantons with $p=q_+=1$ and $q_-=0$, the area must be
 computed separately. One finds $16\pi n a$.}
\begin{equation}
     A = {16\pi n^2 \over p \Omegahat ^2}
     \left(-
     \betahat + \sqrt{\betahat^2 - \Omegahat^2 \over 1 - \Omegahat^2}
     \right) \ .
\end{equation}

In the generic case, then, only two of our ``tests'' have proved to be
nontrivial: 
that $M=\M$, 
and (if we wish to include it)
that $S=A/4G$.
The first succeeds, the second fails.

In figures (\ref{Entropy}) and (\ref{Entropyaft}) the behaviours of
entropy before and and after analytic continuation are presented. 

\section{Open questions and closed ones}

To claim that black hole thermodynamics in the presence of NUT charge
$n$ is in satisfactory shape would be to ignore a whole host of
difficulties that stem from the topological twisting that arises when
$n\not=0$.  One of these difficulties --- not previously appreciated as
far as we know --- is that suitable boundary conditions fail to
exist for all the possible values of the thermodynamic parameters
$\beta$ and $\Omega$.  And even when such boundary conditions do exist,
there is not always an instanton to satisfy them, if we exclude
singularities like ``Misner strings'' and conical points
(e.g. there are no instantons for $p>2$, $q_\pm=1$).
This suggests that new instantons might be found corresponding to these
``unfulfilled'' boundary conditions.  
However, the question is confused by the fact that different
choices of the vector $\xi$ can give rise to the same lattice $\L$.
(See footnote \eight above.)

However, 
if we limit ourselves to the ``Kerr-Bolt'' line element (\ref{line-elt}), 
we can classify all nonsingular instantons.  
We find solutions only for certain 
triples of relatively prime integers $p,q_\pm$.
In particular, the parameter $\Omtilde=\beta\Omega$, 
with $\beta$ and $\Omega$ defined by (\ref{xi-def}), 
is restricted to rational multiples of $2\pi$,
as followed already from the asymptotic analysis.  
These (purely classical) quantization rules are perhaps 
the most surprising outcome of the above analysis. 

The mathematical problem of finding all ``Kerr-Bolt instantons''
is thus largely solved.  The thermodynamic meaning of these solutions,
on the other hand, is still in the dark to a significant extent. 
The obscurities begin with the correct identification of the parameters
$\Omega$ and $\beta$ in terms of the asymptotic metric, and they end
with the unknown reason for the failure of the normal area law for the entropy. 
In between are numerous other unanswered questions:
how to define conserved quantities like energy when the $t=0$
hypersurface is not closed;
how to perform the analytic continuation in $\Omega$ that would
transform $I_E$ into (an approximation to) the seminal thermodynamic
potential $\Psi$;
how to recover a thermodynamic angular momentum $J$
when $\Omtilde$ is apparently not free to vary;
how to interpret a lens space topology at infinity (especially when it
is not a fibration over $S^2$);
etc.

This paper had its origins in the failure of the first law to hold when
one simplistically carries over the $n=0$ formulas (for the Kerr
instanton) to the $n\not=0$ situation.  
We have seen that this failure was misleading, because the
simplistic formulas are incorrect on several counts.  
For example, the correct value of the $d\varphi dt$ integral that
enters into $I_E$ is not $2\pi\beta$, but $16\pi^2 n^2/p$.
When these corrections are made, and when the limitations on $\beta$ and
$\Omtilde$ are taken into account, the various thermodynamic relationships
(including the first law) appear to be satisfied to the extent that they
continue to make sense.

However, the obscurities (some of) which we enumerated above leave our
results in the previous section somewhat in limbo.
Perhaps, a clearer insight into the analytic continuation question could
be had, and could 
improve one's understanding.  Or perhaps the attempt to do black hole
thermodynamics with $n\not=0$ is a case of pushing concepts outside
their valid domain of application.

One way to help decide which is correct would be to study non-rotating
but charged instantons with $n\not=0$, where analogous difficulties
with the naive thermodynamic relationships are known to arise.

Another interesting question is why the equality $S=A/4G$ breaks down
when $n\not=0$.  In light of \cite{sumati}, this failure appears to be
bound up with the failure of a ``Cauchy surface'' to exist, but it would
be nice to have a more quantitative understanding of the discrepancy.

In concluding, we would like to return to our exceptional solution in
order to point out a suggestive analogy it bears to a thermodynamic
system undergoing a first order phase transition
like
the melting of
ice.  In such a situation two phases are present simultaneously and a
new equilibrium parameter arises in addition to the intensive
thermodynamic variables characterizing the ``reservoir'', namely a
parameter whose variation signifies the conversion of one phase into the
other.  For a reservoir characterized by a temperature and an angular
velocity, such a transition will typically occur along a critical line
in the $\beta$-$\Omega$-plane.
Now in our situation the analogous line is given by
$\Omtilde=\beta\Omega=2\pi$ which is 
a hyperbola in the $\beta$-$\Omega$-plane or 
a horizontal line in the $\beta$-$\Omtilde$-plane.  
(In fact, the instantons we have found yield a bounded --- and totally
disconnected --- subset of the plane, and our critical line lies on the
boundary of the region occupied by this subset.)  
Along
this line, there arises, as we have seen, a new parameter that becomes
independent of $n$.  Let us take this parameter to be $a$, varying
between $0$ and $\infty$.  [Regularity of the line-element
(\ref{line-elt}) (in the given coords) restricts $r_0$ to the range
$(n,\infty)$, while $a=r_0-n$ for our solution.]  

In the spirit of our analogy, we could think of $a$ as analogous to the
amount of ice present in a mass of freezing water.  But does some
``substance'' actually disappear when $a=0$?  In a sense, the answer is
yes and the substance is the black hole horizon!  
With $r_0=n+a$,
(\ref{line-elt}) becomes spherically symmetric, 
and the area of the
horizon reduces simply to $4\pi(r_0^2-n^2)=4{\pi}a(a+2n)$.
Thus, $a$ is essentially just measuring the horizon radius in this case.  
When $a\to0$ the
horizon 
shrinks away like a chunk of ice in a warming lake.
We have here, perhaps, one more instance in which the formation or
disappearance of a
black hole bears all the marks of a thermodynamic phase transition.

\bigskip

\noindent{\Large Acknowledgments}

This work was supported by the Natural Sciences and Engineering Research
Council of Canada
and by the National Science Foundation of the United States
under grant PHY-0404646.
Research at Perimeter Institute for Theoretical Physics is supported in
part by the Government of Canada through NSERC and by the Province of
Ontario through MRI.

\section{Appendix 1: Kerr-Bolt-(A)dS spacetimes}

The Lorentzian geometry of Kerr spacetimes with NUT charge and nonzero
cosmological constant is given by the line element%
\begin{equation}
ds^{2}=-\frac{\Delta _{L}(r)}{\Xi _{L}^{2}\rho _{L}^{2}}[dt+(2n\cos \theta
-a\sin ^{2}\theta )d\varphi ]^{2}+\frac{\Theta _{L}(\theta )\sin ^{2}\theta }{%
\Xi _{L}^{2}\rho _{L}^{2}}[adt-(r^{2}+n^{2}+a^{2})d\varphi ]^{2}+\frac{\rho
_{L}^{2}dr^{2}}{\Delta _{L}(r)}+\frac{\rho _{L}^{2}d\theta ^{2}}{\Theta
_{L}(\theta )} \ ,  \label{LKBdS}
\end{equation}%
where%
\begin{equation}
\begin{array}{c}
\rho _{L}^{2}=r^{2}+(n+a\cos \theta )^{2} \ , \\ 
\Delta _{L}(r)=-\frac{r^{2}(r^{2}+6n^{2}+a^{2})}{\ell ^{2}}+r^{2}-2mr-\frac{%
(3n^{2}-\ell ^{2})(a^{2}-n^{2})}{\ell ^{2}} \ , \\ 
\Theta _{L}(\theta )=1+\frac{a\cos \theta (4n+a\cos \theta )}{\ell ^{2}} \ , \\ 
\Xi _{L}=1+\frac{a^{2}}{\ell ^{2}} \ ,%
\end{array}
\label{TNKdSLfuncs}
\end{equation}%
which are exact solutions of the Einstein equations.
The ``event horizons'' of
the spacetime are given by the singularities of the metric function which
are the real roots of $\Delta _{L}(r)=0$. These are determined by the
solutions of the equation%
\begin{equation}
  r_{+}^{4}-r_{+}^{2}(\ell ^{2}-6n^{2}-a^{2})
  +2m\ell^{2}r_{+}+(3n^{2}-\ell^{2})(a^{2}-n^{2})=0  \ .
 \label{TNKDShorizons}
\end{equation}

The ``Euclidean section'' for this class of metrics is given by 
\begin{equation}
ds^{2}=\frac{\Delta _{E}(r)}{\Xi _{E}^{2}\rho _{E}^{2}}[dt-(2n\cos \theta
-a\sin ^{2}\theta )d\varphi ]^{2}+\frac{\Theta _{E}(\theta )\sin ^{2}\theta }{%
\Xi _{E}^{2}\rho _{E}^{2}}[adt-(r^{2}-n^{2}-a^{2})d\varphi ]^{2}+\frac{\rho
_{E}^{2}dr^{2}}{\Delta _{E}(r)}+\frac{\rho _{E}^{2}d\theta ^{2}}{\Theta
_{E}(\theta )} \ . \label{EKBdS}
\end{equation}%
where we have analytically continued the $t$ coordinate, the NUT charge and
the rotation parameter to imaginary values, yielding%
\begin{equation}
\begin{array}{c}
\rho _{E}^{2}=r^{2}-(n+a\cos \theta )^{2} \ ,\\ 
\Delta _{E}(r)=-\frac{r^{2}(r^{2}-6n^{2}-a^{2})}{\ell ^{2}}+r^{2}-2mr-\frac{%
(3n^{2}+\ell ^{2})(a^{2}-n^{2})}{\ell ^{2}} \ , \\ 
\Theta _{E}(\theta )=1-\frac{a\cos \theta (4n+a\cos \theta )}{\ell ^{2}} \ ,\\ 
\Xi _{E}=1-\frac{a^{2}}{\ell ^{2}} \ . %
\end{array}%
\end{equation}

The Euclidean section exists only for values of $r$ such that the function $%
\Delta _{E}(r)$ is positive valued. The horizons are located at the zeros of 
$\Delta _{E}(r)$, which we shall denote by $r=r_{0}$. Moreover, the range of 
$\theta $ depends strongly on the values of the NUT charge $n$, the
rotational parameter $a$ and the cosmological constant $\Lambda =3/\ell ^{2}$%
, taken here to be positive (for the solution with $\Lambda <0$, replace $%
\ell ^{2}$ with $-\ell ^{2}$ in the preceding equations). 
The angular
velocity of the horizon 
in the Lorentzian geometry is given by
\begin{equation}
\Omega _H = \left. -\frac{g_{t\varphi }}{g_{\varphi \varphi }}\right| _{r=r_{0}}=%
\frac{a}{r_{0}^{2}+n^{2}+a^{2}} \ , \label{surface}
\end{equation}%
and so the surface gravity of the cosmological horizon can be calculated to
give%
\begin{equation}
\kappa =\frac{1}{2(r_{0}^{2}+n^{2}+a^{2})\Xi_L }\left. \frac{d\Delta_L }{d r }
\right| _{r=r_{0}} \ ,
\end{equation}%
where the Killing vector 
$\chi ^{\mu }=\zeta ^{\mu }+\Omega_H \psi^{\mu}$
is normal to the horizon surface $r=r_{0}$.




We first note that there are no pure NUT solutions for nonzero values of $a$. 
We demonstrate this as follows.\footnote
{It also follows purely topologically from the fact that the surfaces
 of constant $r$ are not homeomorphic to $S^3$.}
Since $\psi ^{\mu }$\ is a Killing vector,
for any constant $\varphi $ -slice near the horizon, additional conical
singularities will be introduced in the $(t,r)$\ section unless the period
of $t$\ is $\Delta t=\frac{2\pi }{\left| \kappa \right| }$. Furthermore,
there are string-like singularities along the $\theta =0$ and $\theta =\pi $
axes for general values of the parameters. \ These can be removed by making
distinct shifts of the coordinate $t$ in the $\varphi $ direction near each of
these locations. These must be geometrically compatible 
\cite{stressten},
yielding the requirement that the period of $t$ should be 
$\Delta t=4n\Delta \phi=\frac{16\pi n}{q_++q_-}$. 
Demanding the absence of both
conical and Dirac-Misner singularities, we get the relation 
\begin{equation}
  \frac{k}{\kappa}=\frac{8n}{q_++q_-} \ , 
  \label{betaeq8Pin}
\end{equation}%
where $k$ is any non-zero positive integer. Demanding the existence of a
pure NUT solution at $r=r_{0}$ is equivalent to the requirement that that
the area of the surface of the fixed point set of the Killing vector $%
\partial /\partial t,$%
\begin{equation}
 A=\frac{2\pi}{\Xi_{E}} 
   \{ \alpha \sqrt{\Delta_E(r_{0})} + 2(r_{0}^{2} - n^2-a^2) \}  \ ,
  \label{surfacefixed}
\end{equation}%
vanish, 
where $\alpha =\int_{0}^{\pi }\frac{2n\cos \theta -A\sin ^{2}\theta}
{\sqrt{\Theta _{E}(\theta )}}d\theta $. 
In other words, this surface is of
zero dimension. This can only occur special values of the mass parameter.
However if we select for this parameter we find an inconsistency with the
relation (\ref{betaeq8Pin}), which must hold for the spacetime with NUT
charge.

Hence we conclude that the only spacetimes with NUT charge and rotation is
Taub-Bolt-Kerr-(A)dS spacetimes, where the term ``bolt'' refers to the fact
that the dimensionality of the the fixed point set of $\partial /\partial t$
is two.

\bigskip

\section{Appendix 2: Commensurability Conditions}

{\it 
Let $\vec{u},\vec{v},\vec{w}$ be three vectors such that
\begin{equation}
  p\vec{u} + q\vec{v} + r\vec{w} = 0 \ ,
 \label{eq}
\end{equation}
with $p,q,r \in \Z$.
Without loss of generality, $gcd(p,q,r)=1$. 
Assume that
$\vec{u},\vec{v},\vec{w}$ are minimal members of 
the lattice $\L=\span_{\Z}\braces{\vec{u},\vec{v},\vec{w}}$.
Then $p,q$ and $r$ are pairwise relatively prime. 
}

\medskip

To prove this, assume the opposite. 
Say for example that $p$
and $q$ have a common factor $\phi$: $p=a\phi , q=b\phi $; then
$\frac{r}{\phi}\notin \Z$.  From equation (\ref{eq}), we conclude $\frac{r}{\phi} \vec{w} \in \L$ because
$a,b \in \Z$. 
On the other hand since
$\frac{r}{\phi} \notin \Z$, 
we can set 
$\frac{r}{\phi}=c+\lambda$ where $c\in \Z$ and 
$0 < \lambda < 1$. 
From $\frac{r}{\phi} \vec{w} \in \L$ and $c\vec{w} \in \L$, 
it follows that
$\frac{r}{\phi} \vec{w}-c\vec{w} \in \L$,
whence $\lambda \vec{w} \in \L$
contradicting our assumption that $\vec{w}$ was minimal.

{\it 
Conversely, let $\vec{u},\vec{v}$ and $\vec{w}$  three vectors 
not all parallel to each other and assume that
they satisfy
(\ref{eq}) with $p,q,r$ pairwise relatively prime.
Then $\vec{u},\vec{v}$ and $\vec{w}$
are minimal in $\L$.  }
\medskip

Let us  prove for example that $\vec{v}$ is minimal.
Either $\vec{u}$ or $\vec{w}$ is independent of $\vec{v}$, say it is the
former. 
Now take any general vector $\vec{x}$ in $\L$ as
$\vec{x}=a\vec{u}+b\vec{v}+c\vec{w}=(a-\frac{pc}{r})\vec{u}+(
b-\frac{qc}{r})\vec{v}$, where $a,b,c \in \Z$.
If this vector is proportional to $\vec{v}$ then necessarily
$r \mid pc$ because $\vec{u}$ and $\vec{v}$ are linearly independent.
Since $r$ and $p$ are relatively prime, 
$r \mid c$, 
whence $q{c}/{r} \in \Z$, 
whence $\vec{x}$ is an integer multiple of $\vec{v}$.
This shows $\vec{x}$ is not a submultiple of $\vec{v}$. 
Hence $\vec{v}$ is minimal.

\bigskip

\section{Appendix 3: Some information on Lens spaces}

A lens space is a 3-manifold that can be obtained as the quotient of the
3-sphere $S^3$ by the action of a cyclic group 
$\Z_p = \Z \, {\rm mod } \, p = \Z/p\Z$.  It can
also be obtained by sewing together two solid tori, but the 
presentation as a quotient is more useful in the present context.

Let $S^3$ be represented as the unit sphere in $\Reals^4$, or rather
as the set of pairs of complex numbers $\zeta=(\zeta^1,\zeta^2)$ such
that $|\zeta^1|^2+|\zeta^2|^2=1$.  If we think of $\zeta^1$  and
$\zeta^2$ as the components of a spinor $\zeta$, then $S^3$ is simply the
set of unit spinors, i.e. spinors such that $\zeta^\dagger\zeta=1$.  
We can relate this type of coordinatization to ``polar coordinates''
$r$, $\theta$, $\varphi$, $\psi$ by setting
\begin{equation}
    \zeta = \sqrt{r} e^{i\psi/2} \,
    (\cos(\theta/2) e^{-i\varphi/2}, \, \sin(\theta/2)e^{i\varphi/2}) \ .
      \label{zetacoords}
\end{equation}
Here $\th$ ranges from 0 to $2\pi$ and an elementary domain for 
$\varphi$ and $\psi$ can be taken to be $[0,2\pi]\times[0,4\pi]$, or
more symmetrically, the square in the $\varphi$-$\psi$-plane spanned by
the vectors,
$\eta_\pm = 2\pi (\partial /\partial \psi \pm \partial /\partial \varphi)$,
introduced in (\ref{etaplus}), (\ref{etaminus}).

Now let $q_1, \, q_2 \in \Z_p$ be two integers taken modulo $p$, 
and let $\1^x$
denote the exponential $\exp(2\pi i x)$.
The transformation $g$ defined by
\begin{equation}
   \zeta^j \mapsto \1^{q_j/p} \zeta^j  \quad (j=1,2)  \ ,
\end{equation}
generates the action of a cyclic group $G$ on $S^3$.  
In terms of the angles, $\varphi$ and $\psi$, it is clear from 
(\ref{zetacoords}) that $g$ is equivalent to translation by the vector 
\begin{equation}
   \xi = {q_2\over p} \eta_{+} + {q_1\over p} \eta_{-}   \ .
\end{equation}
Let $L(p;q_1,q_2)$ be the quotient of the 3-sphere by this action:
\begin{equation}
     L(p;q_1,q_2) = S^3 / G \ .
\end{equation}
Plainly, $G$ is a cyclic group of order at most $p$, since $g^p=1$.
Without modifying $g$ we can clearly divide out any common factor in 
$p$, $q_1$, $q_2$; then $\gcd(p,q_1,q_2)=1$.  
Having done so, we claim further
that both $q_1$ and $q_2$ must be relatively prime to $p$
(that is, the phases $\1^{q_j/p}$ must be primitive $p$th roots of unity.);
for in the contrary case
we would have (say)
$r|p$, $r|q_1$, whence
\begin{equation}
 g^{p/r}=\diag(\1^{{p\over r}{q_1\over p}},\1^{{p\over r}{q_2\over p}})
        =\diag(\1^{q_1\over r},\1^{q_2\over r})
        =\diag(1,\1^{q_2\over r}) \ ,
\end{equation}
since $r|q_1$.  But this transformation has the ``north
pole'' $\zeta=(1,0)$ as a fixed point, while it is not the identity in
$G$ because $r$ cannot divide $q_2$ if it divides $p$ and $q_1$.  Thus, 
$S^3/G$ would not be a manifold.  Conversely, if $q_1$
and $q_2$ are both relatively prime to $p$ then $G$ has no fixed points
and then $S^3/G$  is a manfold.  
We thus obtain a lens space 
$L(p;q_1,q_2)$
for every
triple of integers $(p,q_1,q_2)$, where the $q_j$ are taken modulo
$p$ and must be prime to it. 
Comparing with (\ref{xidef}) then shows that
the instanton constructed in Section 4 with parameters $p,q_\pm$, is
topologically the lens space $L(p;q_-,q_+)$.

In fact, we can go further and require the $q_j$ to be relatively prime
to each other as well.  
If they were not, they would have a common
factor $r$, so without changing $G$, we could replace $q_j$ by
$q_j/r$, which modulo $p$ coincides with
$(1/r)q_j$, where $(1/r)$ is the reciprocal of $r$ in $\Z_p$.  (This
reciprocal exists since $r$ and $p$ are relatively prime.)
It follows that we exhaust the lens spaces by letting $q_1$ and $q_2$
run over all pairs of integers in $\Z_p$ that are relatively prime to
each other and to $p$.   In fact, the same reasoning demonstrates that
we leave nothing out by limiting ourselves to the form $L(p;q,1)$.
This in turn allows a simpler notation $L(p,q)$ for lens spaces,
related to the earlier notation by
\begin{equation}
   L(p;q_1,q_2) =  L(p;q_1/q_2,1) \ideq L(p,q_1/q_2)   \ ,
\end{equation}
where, of course, the division is done in $\Z_p$.
(This equation breaks down when $p=1$ because $0$ and $1$ are not
distinct elements in $\Z_1$ and $0/0$ is not defined a priori.   
For this special case, the natural 
convention to adopt is $0/0=0$ so that $L(1;0,0)$ and $L(1,0)$ both
denote the 3-sphere $S^3$.)   

Finally, we can ask which of the $L(p;q_1,q_2)$ are homeomorphic to each
other.  
First of all, we can observe that $L(p;q_1,q_2)$ depends only on
the {\it ratio} $q = q_1/q_2$, 
as we have already seen in effect, and
as follows explicitly from the simple calculation,
$L(p,q_1,q_2)=L(p;rq_1,rq_2)=L(p;q,1)=L(p,q)$, 
where $r=1/q_2$,
the reciprocal of $q_2$ in $\Z_p$. 
Now by exchanging $\zeta^1$ with $\zeta^2$
(a diffeomorphism of $S^3$),
we learn that $L(p,q)=L(p,1/q)$.
Similarly we have the
equivalence $L(p,q) = L(p,-q)$, by complex conjugation of $\zeta^2$ (say).
Combining these yields the set of equivalences,
\begin{equation}
        L(p,q) = L(p,\pm q^{\pm 1})  \ .
\end{equation}
According to \cite{lens}, there are no other equivalences among the
$L(p,q)$.  That is, $L(p,q)$ is homeomorphic to $L(p',q')$ if and only if 
$p'=p$ and $q'=\pm q^{\pm 1}$, where the latter equality is modulo $p$,
i.e. $q$ and $q'$ are regarded as elements of $\Z_p$. 
If $L(p,q)$ is treated as an oriented manifold, then the second
equality should be replaced by simply $q'=q^{\pm 1}$.

The lens space $L(1;1,1)$ is simply $S^3$ itself, and the above
vector $\xi$ in this case generates (as a continuous symmetry, not a
discrete one) the Hopf fibration of $S^3$ over $S^2$.  Similarly
the 
$L(p;1,1)=L(p,1)$ 
are circle bundles over $S^2$ with $p$ twists --- monopoles
of charge $p$ in the Kaluza-Klein interpretation.  Since there are no
other circle bundles over $S^2$, it follows that all of the other lens
spaces are something else; and the corresponding instantons cannot 
possess a full $S^2${}'s worth of directions at infinity.

\end{document}